\documentclass{appolb}
\usepackage{epsfig}
\usepackage{amssymb}
\begin{document}
\pagestyle{plain}
\newcount\eLiNe\eLiNe=\inputlineno\advance\eLiNe by -1
\title{Detecting subtle effects of persistence in the stock market
dynamics
}
\author{R.Rak$^1$, S.~Dro\.zd\.z$^{1,2}$, J.~Kwapie\'n$^2$, P.~O\'swi\c ecimka$^2$
\address{$^1$Institute of Physics, University of Rzesz\'ow, PL--35-310
Rzesz\'ow, Poland\\
$^2$Institute of Nuclear Physics, Polish Academy of Sciences, \\
PL--31-342 Krak\'ow, Poland}} \maketitle

\begin{abstract}
The conventional formal tool to detect effects of the financial
persistence is in terms of the Hurst exponent. A typical
corresponding result is that its value comes out close to $0.5$,
as characteristic for geometric Brownian motion, with at most
small departures from this value in either direction depending on
the market and on the time scales involved. We study the high
frequency price changes on the American and on the German stock
markets. For both corresponding indices, the Dow Jones and the DAX
respectively, the Hurst exponent analysis results in values close
to $0.5$. However, by decomposing the market dynamics into pairs
of steps such that an elementary move up (down) is followed by
another move up (down) and explicitly counting the resulting
conditional probabilities we find values typically close to
$60\%$. This effect of persistence is particularly visible on the
short time scales ranging from 1 up to 3 minutes, decreasing
gradually to $50\%$ and even significantly below this value on the
larger time scales. We also detect some asymmetry in persistence
related to the moves up and down, respectively. This indicates a
subtle nature of the financial persistence whose characteristics
escape detection within the conventional Hurst exponent formalism.
\end{abstract}

\section{Introduction}
The financial dynamics results in fluctuations whose nature is, as
pointed out by Bachelier~\cite{Bach} already in 1900, of the
Brownian character. By now we know that it is much more complex
and fascinating than just the ordinary Brownian motion. Already
the distribution of stock market returns is far from being
Gaussian and at the short time scales large fluctuations develop
heavy power law asymptotics with an exponent
$\alpha=3$~\cite{Gop}, well outside the Levy stable
regime~\cite{Dro1}, however. The autocorrelation function of
returns drops down very quickly and after a few minutes it reaches
the noise level. At the same time however the volatility
autocorrelation function decays very slowly with time~\cite{Dro1},
largely according to the power law, and remains positive for many
months. On a more advanced level of global quantification, the
financial dynamics appears to be describable in terms of
multifractality both in the transaction-to-transaction price
increments and in the inter-trade waiting times~\cite{Osw}. This
indicates a hierarchically convoluted self-similar organization of
the market dynamics. One related issue is an effect of
persistence. Its commonly adopted measure - the Hurst exponent -
is the mode of each multifractal spectrum. The Hurst exponent,
however, is a global measure while the effects of persistence may
in principle depend on the market phase. Below we address this
issue using the high-frequency records (years 1998-99) of the two
from among the world leading stock market indices, the Dow Jones
Industrial Average (DJIA) for the United States and the Deutsche
Aktienindex (DAX) for Germany.

\section{Conventional methods}

There exist two commonly accepted and best-known methods to
evaluate the long-range dependences in the statistical series. The
older one is the so-called rescaled range or R/S
analysis~\cite{Man}. This method originates from previous
hydrological analysis of Hurst~\cite{Hur} and allows to calculate
the self-similarity parameter $H$. A drawback of this method
however is that it may spuriously detect some apparent long-range
correlations that result from non-stationarity. A method that
avoids such artifacts is the Detrended Fluctuation Analysis
(DFA)~\cite{Peng}. In this method one divides a time series
$g(t_i)$ of length $N$ $(i=1,...N)$ into $M$ disjoint segments
$\nu$ of length $n$ and calculates the signal profile
\begin{equation}
Y_{\nu} (i)=\sum_{k=1}^i (g(k)-\langle g \rangle),~~~~i=1,...,N
\label{sp}
\end{equation}
where $\langle ... \rangle$ denotes the mean. For each segment
$\nu$ the local trend is then estimated by least-squares fitting
the straight line (in general a polynomial) ${\tilde Y}_{\nu}(i)$
and the corresponding variance
\begin{equation}
F^2(\nu, n) = {1 \over n} \sum_{j=1}^n \{ Y[(\nu - 1)n + j] -
{\tilde Y}(j)\}. \label{var}
\end{equation}
Finally, one calculates the mean value of the root mean square
fluctuations over all the segments $\nu$:
\begin{equation}
{\bar F}(n) = {1\over M} \sum_{\nu = 1}^M F (\nu, n) \label{msf}
\end{equation}
The power-law scaling of the form
\begin{equation}
{\bar F}(n) \sim n^H \label{pl}
\end{equation}
indicates self-similarity and is considered to provide a measure
of persistence. If the process is white noise then $H=0.5$. If the
process is persistent then $H > 0.5$; if it is anti-persistent
then $H < 0.5$.

The above procedure applied to the returns
\begin{equation}
g(t) = \ln P(t + \Delta t) - \ln P(t), \label{ret}
\end{equation}
where $P(t)$ represents the price time series, results in numbers
as listed in the last column of Table 1 for $\Delta t$ ranging
from 1 min up to 30 min. In addition to the DAX and the DJIA this
Table includes also what for brevity we here call Nasdaq30 and
what for the purpose of this work is constructed as a simple sum
of the prices of 30 high-capitalization companies belonging to the
Nasdaq Composite basket. As one can see from the Table 1,
typically the so-calculated Hurst exponents $H$ point to a trace
of anti-persistence but in fact they do not deviate much from 0.5,
especially that an error involved in estimation equals about
$0.4\%$ for $\Delta t = 1$ min and increases up to $1.5\%$ for
$\Delta t = 30$ min due to an effective shortening of the series.

Still this result does not eliminate a possibility that there
exist some more local effect of persistence that simply average
out when estimated from the longer time intervals. In fact, some
proposals to calculate the local counterparts of $H$, based on
variants of DFA, are already present in the
literature~\cite{Gre,Car} and point to such effects indeed. The
accuracy of the related methods is however not yet well
established. Furthermore, observations and experience prompt a
possibility that the financial persistence may happen to occur
asymmetrically, i.e., a move up may be followed by another move up
more often than a move down by another move down, or vice versa.
Such effects may carry a very valuable information about the
dynamics but remain indistinguishable within the conventional
methods and unexplored so far. In order therefore to explore a
possibility and character of such effects we return to the very
definition of persistence.

\section{Measuring persistence by conditional probabilities}

Given a time series of price returns $g(t_i)$, where $t_i$ denotes
the consecutive equidistant moments of time, to each $i$ we assign
+1 if $g(t_i)$ is positive (price goes up), -1 if it is negative
(price goes down) and 0 if it happens to be 0 (price remains
unchanged). We then explicitly count the number $N_{\alpha,\beta}$
of all the neighboring pairs $\{ g(t_i), g(t_{i+1})\}$ of the type
$\alpha, \beta = \{-1, 0, +1 \}$ for fixed values of $\alpha$ and
$\beta$ and do so for all the nine combinations of different
$\alpha$ and $\beta$. Finally, we calculate
\begin{equation}
p_{\alpha, \beta} = N_{\alpha, \beta} / \sum_{\beta'=-1, 0, +1}
N_{\alpha, \beta'},\label{cp}
\end{equation}
which corresponds to a conditional probability that a return of
the type $\beta$ is preceded by a return of the type $\alpha$.
This procedure can of course be performed on any time scale
$\Delta t = t_{i+1} - t_i$.

Six combinations of $p_{\alpha, \beta}$ corresponding to $\alpha =
\pm 1$ and to all the three possible values of $\beta$ are listed
in Table 1 for several values of $\Delta t$ starting from 1 up to
30 min.\\
\begin{center}
\begin{table}
\begin{footnotesize}
\begin{tabular}
{|c|c|c|c|c|c|c|c|} \hline \bf Index/Scale & $\bf P_{11}$ & $\bf
P_{1-1}$ & $\bf P_{10}$ & $\bf P_{-1-1}$
& $\bf P_{-11}$ & $\bf P_{-10}$ & \bf {Hurst exp.} \\
\hline \bf DAX/1min & 0.567 & 0.427 & 0.005 & 0.562 & 0.431 &
0.005 & 0.493 \\ \hline \bf DAX/2min & 0.568 & 0.428 & 0.002 &
0.561 & 0.435 & 0.003 & 0.496 \\ \hline \bf DAX/3min & 0.554&
0.444 & 0.0002 & 0.548 & 0.449 &  0.0014 & 0.497 \\ \hline \bf
DAX/4min & 0.539 & 0.459 & 0.001 & 0.529 & 0.469 & 0.001 & 0.498
\\ \hline
\bf DAX/5min & 0.528 & 0.470 & 0.0006 & 0.514 & 0.484 & 0.0008 &
0.5 \\ \hline \bf DAX/10min & 0.493 & 0.507 & 0.0002 & 0.476 &
0.522 & 0.0006 & 0.498 \\
\hline \bf DAX/15min & 0.483 & 0.515 & 0.0004 & 0.460 & 0.538 &
0.0009 & 0.498 \\ \hline \bf DAX/30min & 0.483 & 0.516 & 0.0002 &
0.459 & 0.540 & 0.0002 & 0.495\\ \hline
\bf DJIA/1min& 0.558& 0.399& 0.042& 0.558& 0.398& 0.043&  0.502\\
\hline \bf DJIA/2min& 0.555& 0.416& 0.027& 0.561& 0.413&  0.026& 0.499\\
\hline \bf DJIA/3min& 0.526& 0.452& 0.021& 0.531&  0.449& 0.019& 0.498\\
\hline \bf DJIA/4min& 0.504& 0.479& 0.016&  0.504& 0.478& 0.016&
0.498\\ \hline \bf DJIA/5min& 0.498&  0.487&  0.013& 0.497& 0.488&
0.014& 0.495\\ \hline \bf DJIA/10min& 0.497&  0.491& 0.011& 0.498&
0.493& 0.008& 0.491\\ \hline \bf DJIA/15min& 0.502& 0.491& 0.006&
0.487& 0.504& 0.007&  0.491\\ \hline \bf DJIA/30min& 0.506& 0.487&
0.0066& 0.472& 0.521&  0.0068& 0.491\\ \hline \bf NQ30/1min&
0.539& 0.454& 0.006&  0.54&   0.455&  0.005& 0.5\\ \hline \bf
NQ30/2min& 0.547& 0.449&  0.003&  0.546& 0.45& 0.003& 0.499\\
\hline \bf NQ30/3min&  0.539&  0.458& 0.003& 0.529& 0.468& 0.003&
0.501\\ \hline \bf NQ30/4min& 0.532& 0.464& 0.002& 0.518& 0.478&
0.002&  0.499\\ \hline \bf NQ30/5min& 0.53& 0.467& 0.002& 0.515&
0.481&  0.002& 0.501\\ \hline
\bf NQ30/10min&0.538& 0.46& 0.001& 0.511&  0.487& 0.0006& 0.502\\
\hline \bf NQ30/15min&0.526&  0.472&  0.001&  0.497&  0.5& 0.001& 0.5\\
\hline
\end{tabular}
\caption{Several combinations of the conditional probabilities
$p_{\alpha, \beta}$ as defined by Eq.~\ref{cp} for the DAX, DJIA
and for the basket of the largest Nasdaq {\bf (NQ30)} companies
returns on a sequence of different time scales $\Delta t$ ranging
from 1 up to 30 min. The last column lists the corresponding Hurst
exponents. The high-frequency price changes analysed here cover
the time period from 01.12.1997 until 31.12.1999.}
\end{footnotesize}

\end{table}

\end{center}

Quite interestingly - and somewhat unexpectedly in view of the
corresponding values of the Hurst exponents (last column in Table
1) that are very close to 0.5 like for the white noise - both the
DAX and the DJIA show significant effects of persistence on the
small time scales. A move up (down) is followed by another move up
(down) significantly more often than by a move in opposite
direction. For $\Delta t$ larger than 5-10 min we observe a
crossover: the fluctuations become anti-persistent and, what is
particularly interesting, this effect is visibly asymmetric
towards moves down as a systematically observed relation
$p_{-1,-1}$ and $p_{+1,+1}$ indicates. For the basket of the
Nasdaq stocks this crossover also takes place, though on the
somewhat larger time scales. Quite interestingly - and somewhat
unexpectedly in view of the corresponding values of the Hurst
exponents (last column in Table 1) that are very close to 0.5 like
for the white noise - both the DAX and the DJIA show significant
effects of persistence on the small time scales. A move up (down)
is followed by another move up (down) significantly more often
than by a move in opposite direction. For $\Delta t$ larger than
5-10 min we observe a crossover: the fluctuations become
anti-persistent and, what is particularly interesting, this effect
is visibly asymmetric towards moves down as a systematically
observed relation $p_{-1,-1}$ and $p_{+1,+1}$ indicates. For the
basket of the Nasdaq stocks this crossover also takes place,
though on the somewhat larger time scales.

The time period (01.12.1997 - 31.12.1999) of the stock market
variability studied here displays a richness of phases. In the
first half of this period it for instance includes a spectacular
draw up (DAX more than $50\%$) followed by an even faster draw
down to the original level. Previous study~\cite{Dro2} based on
the correlation matrix formalism provides a serious indication
that the dynamics of long-term stock market increases is more
competitive and less collective (as far as correlations among the
individual stocks forming an index is concerned) than during
long-term decreases.

It is thus interesting to inspect if and how our indicators of
persistence correlate with the different phases of the market
dynamics. Two principal such coefficients (rectangles),
$p_{+1,+1}$ and $p_{-1,-1}$, calculated for $\Delta t =1$ min over
one trading day time intervals for all the consecutive days
covering our 01.12.1997 - 31.12.1999 time period, versus the
corresponding DAX changes are shown in Fig.~1. The correlation is
visible indeed.

\begin{figure}[!ht]
\epsfxsize 12.5cm \hspace{-0.5cm} \epsffile{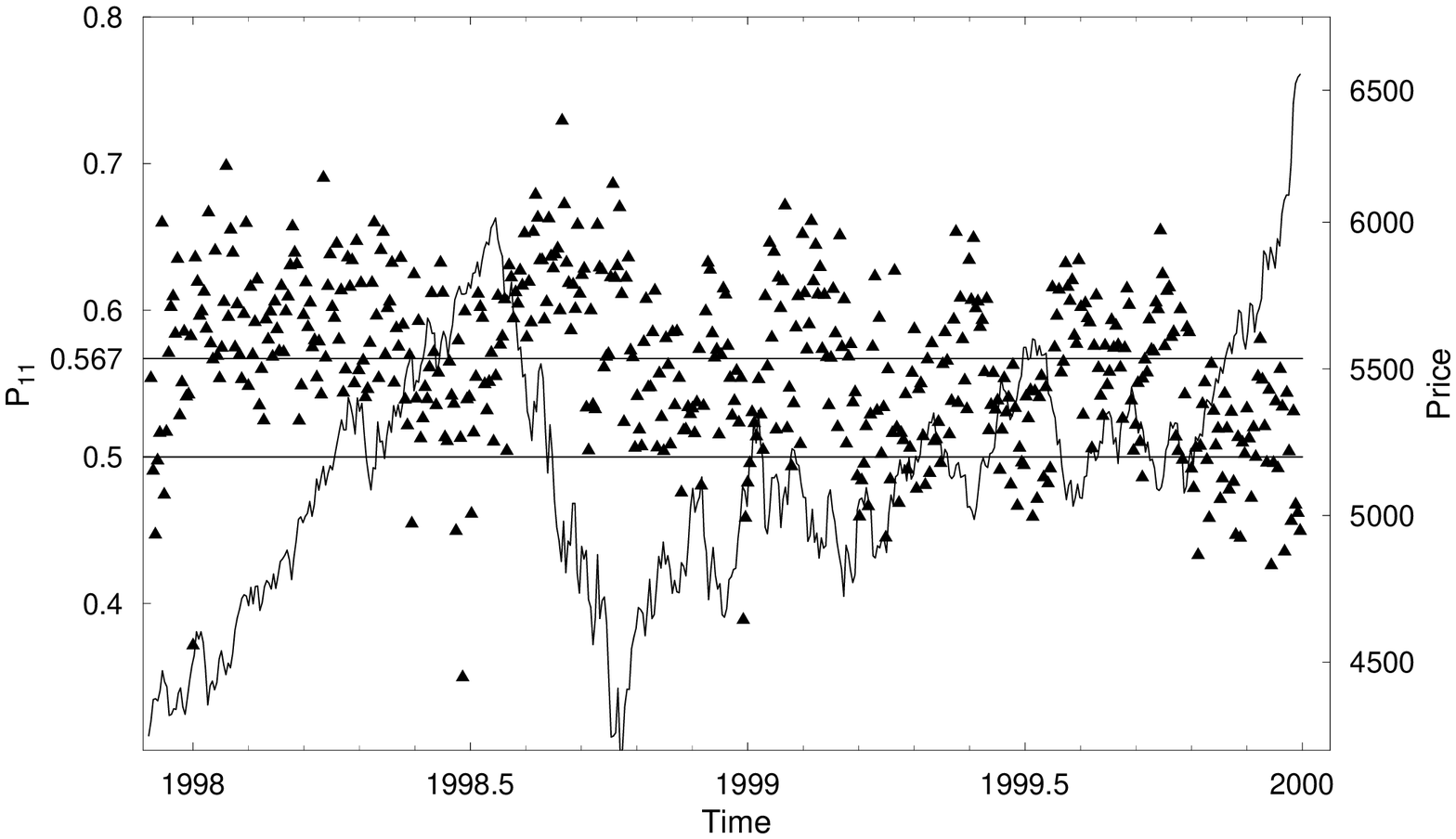}
\vspace{0.0cm} \epsfxsize 12.5cm \hspace{-0.5cm}
\epsffile{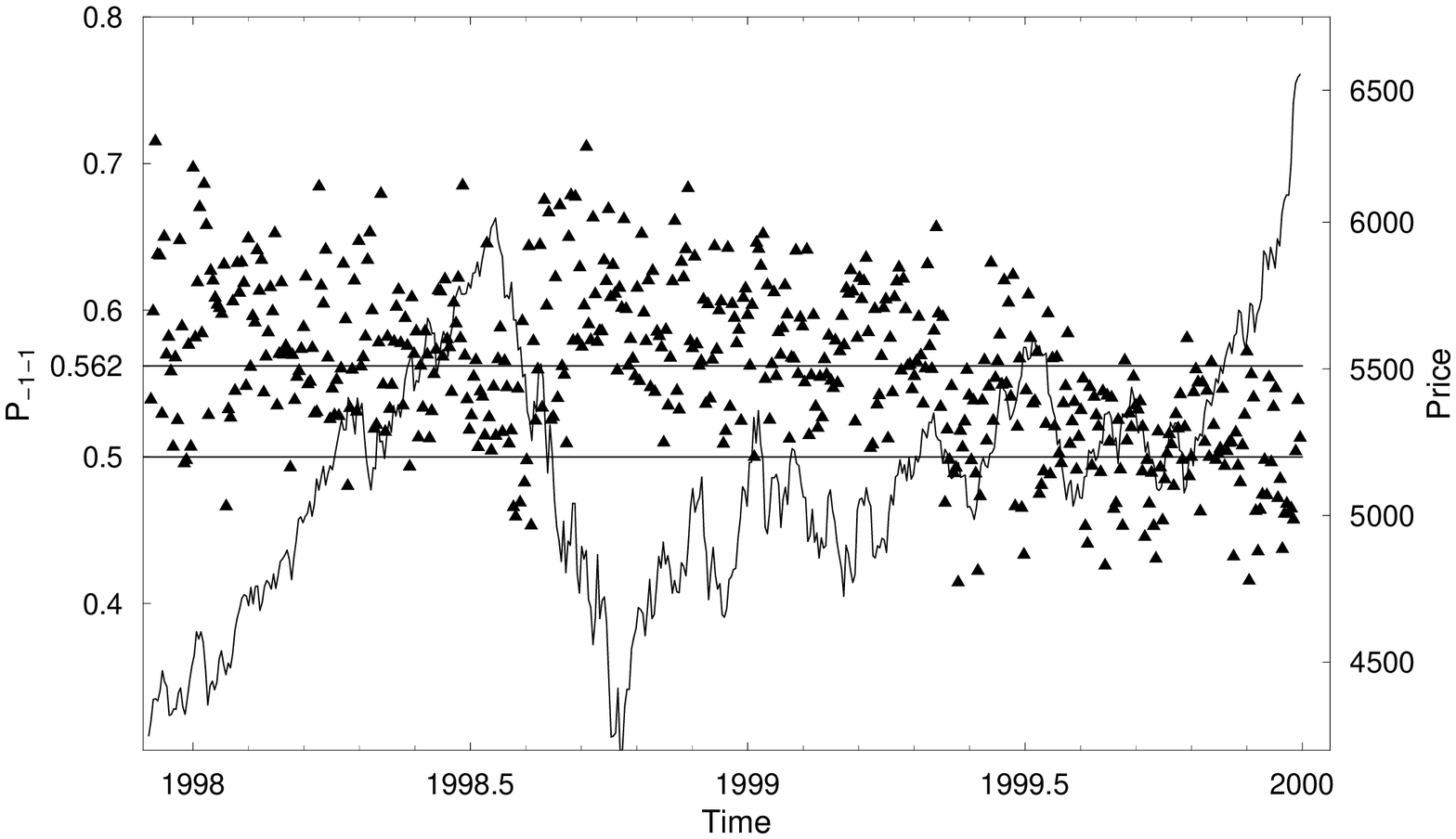}
\begin{footnotesize}
\caption{ The DAX time dependence over the period 01.12.1997 -
31.12.1999. The black triangles ({\scriptsize $\blacktriangle$})
in panels (a) and (b) correspond to $p_{+1,+1}$ and $p_{-1,-1}$,
respectively, calculated separately for the consecutive one
trading day time intervals from $\Delta t=1$ min returns.}
\label{fig1}
\end{footnotesize}
\end{figure}

The long-term increases systematically lead to a decrease of
persistence both in $p_{+1,+1}$ and in $p_{-1,-1}$. A sharp
increase of the DAX in the end of the period here analyzed pulls
these two coefficients even below 0.5. On the contrary, the
decreases are seen to be lifting our persistency coefficients up
to as high as $\sim 0.7$. Even more, changes of the trend in
$p_{+1,+1}$ and $p_{-1,-1}$ are somewhat shifted in phase relative
to each other which is another manifestation of asymmetry in
persistence. That all such effects may reflect a general logic of
the stock market dynamics can be seen also from Fig.~2 which
displays the same quantities for the DJIA in the same period of
time and qualitatively analogous correlations can be postulated.
\begin{figure}[!ht]
\epsfxsize 12.5cm \hspace{-0.5cm} \epsffile{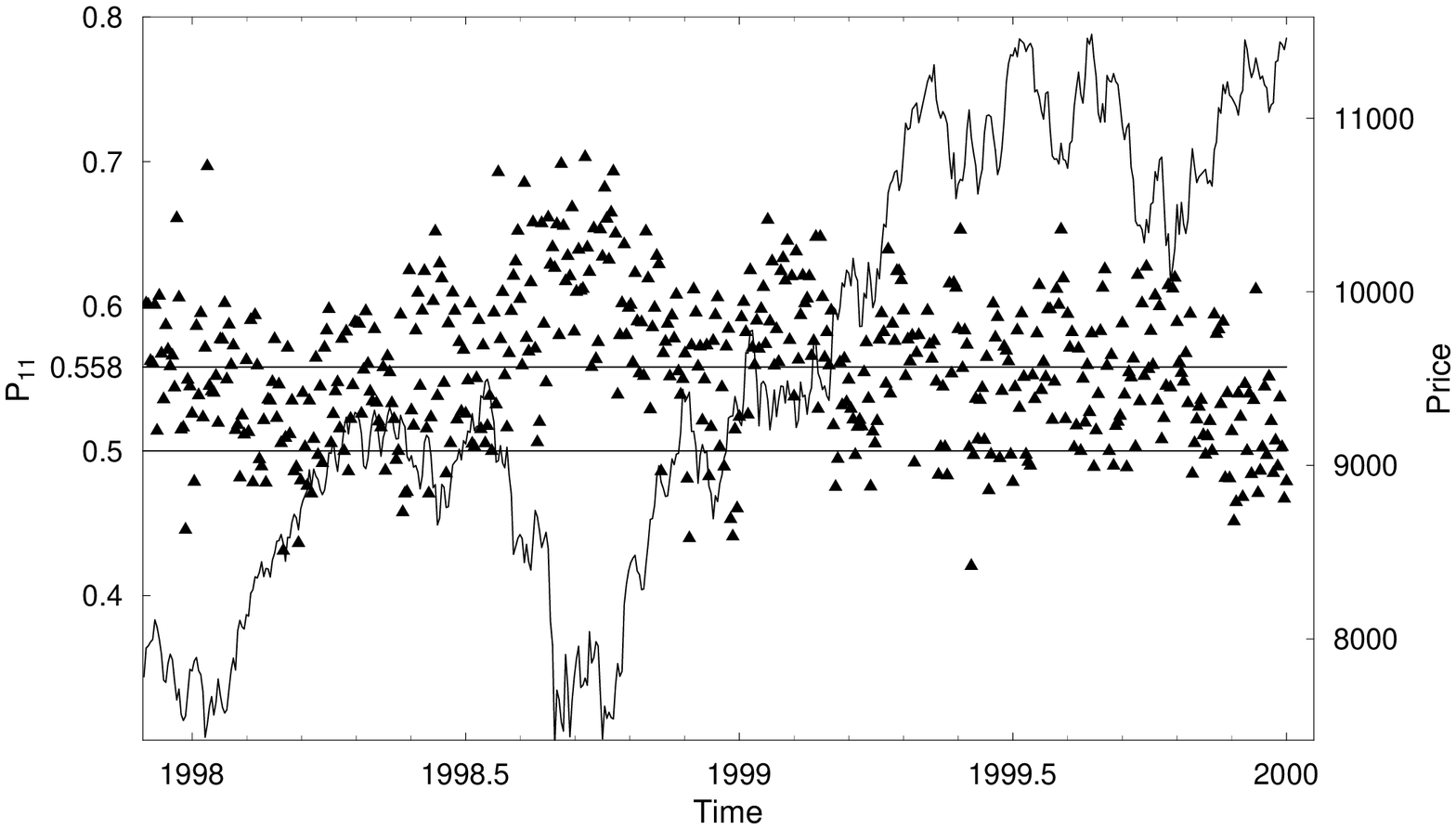}
\vspace{0.0cm} \epsfxsize 12.5cm \hspace{-0.5cm}
\epsffile{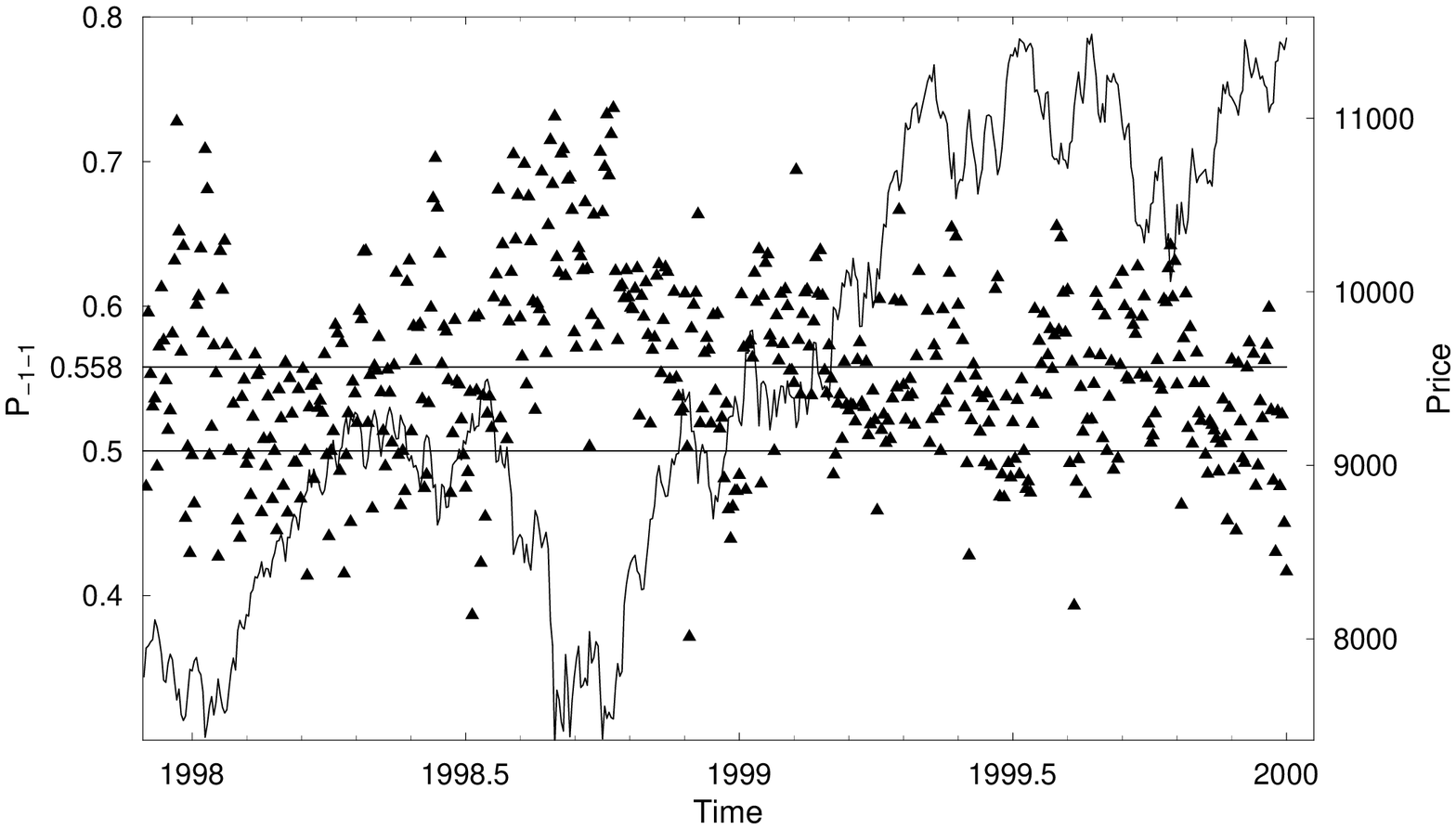}
\begin{footnotesize}
\caption{ Same as Fig.~1 but for the Dow Jones Industrial Average
(DJIA).} \label{fig:fig2}
\end{footnotesize}
\end{figure}

More focus on this last issue is given in Figs.~3 and 4 and the
corresponding Tables 2 and 3. Two sizable periods of the global
market increases and decreases, respectively, both for the DAX and
for the DJIA, are here extracted and the corresponding conditional
probability coefficients $p_{\alpha, \beta}$ calculated for those
periods separately. Again one sees that the related fluctuations
are persistent and that these persistency effects are even
stronger during decreases. Furthermore, even during the same
market phase (either global increase or decrease) the asymmetry in
persistence between the moves up and down may occur.
\begin{figure}[!ht]
\hspace{-0.15cm} \epsfxsize 12.5cm
\epsffile{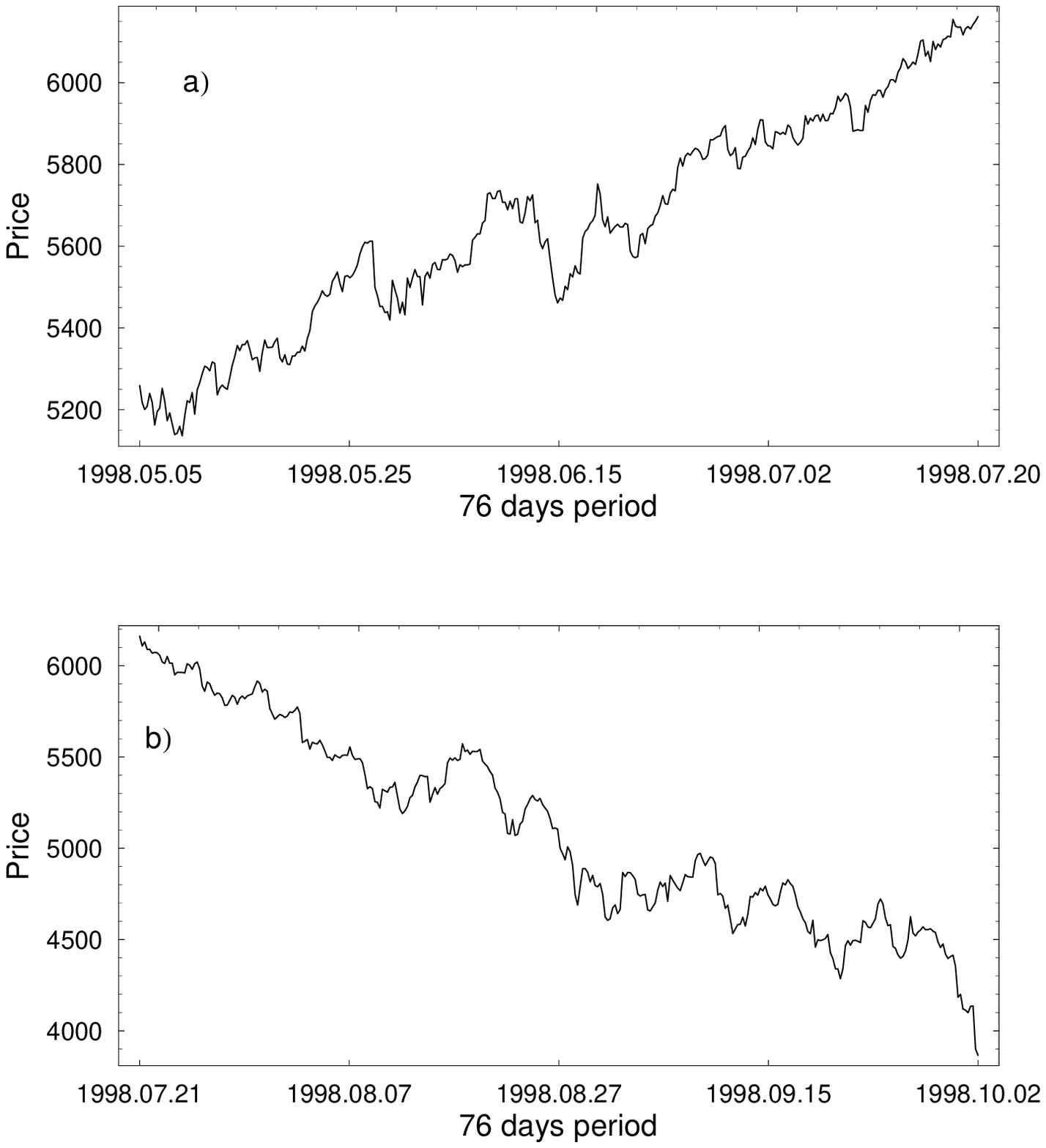}
\begin{footnotesize}
\begin{center}
\begin{tabular}{|c|c|c|c|c|c|c|c|}
\hline \bf Data& $\bf P_{11}$& $\bf P_{1-1}$& $\bf P_{10}$& $\bf
P_{-1-1}$& $\bf P_{-11}$& $\bf P_{-10}$& \bf{Hurst exp.}\\
\hline \bf DAX(increase) &0.562  &0.433  &0.004  &0.57   &0.424  &0.006 &0.491\\
\hline \bf DAX(decrease) &0.615  &0.382  &0.003  &0.585  &0.412
&0.002  &0.504\\ \hline
\end{tabular}
\caption {DAX during its long-term global increase in the period
05.05.1998 - 20.07.1998 (a) and during its long-term global
decrease in the period 21.07.1998 - 02.10.1998 (b). These both
periods correspond to the same number (76) of trading days. Tab.2.
Conditional probabilities $p_{\alpha, \beta}$ and the Hurst
exponents for the DAX $\Delta t=1$ min changes corresponding to
the time periods as in Fig.3.a (DAX/increase) and as in Fig.3.b
(DAX/decrease), respectively.}
\end{center}
\end{footnotesize}
\end{figure}

\begin{figure}[!ht]
\hspace{-0.15cm} \epsfxsize 12.5cm
\epsffile{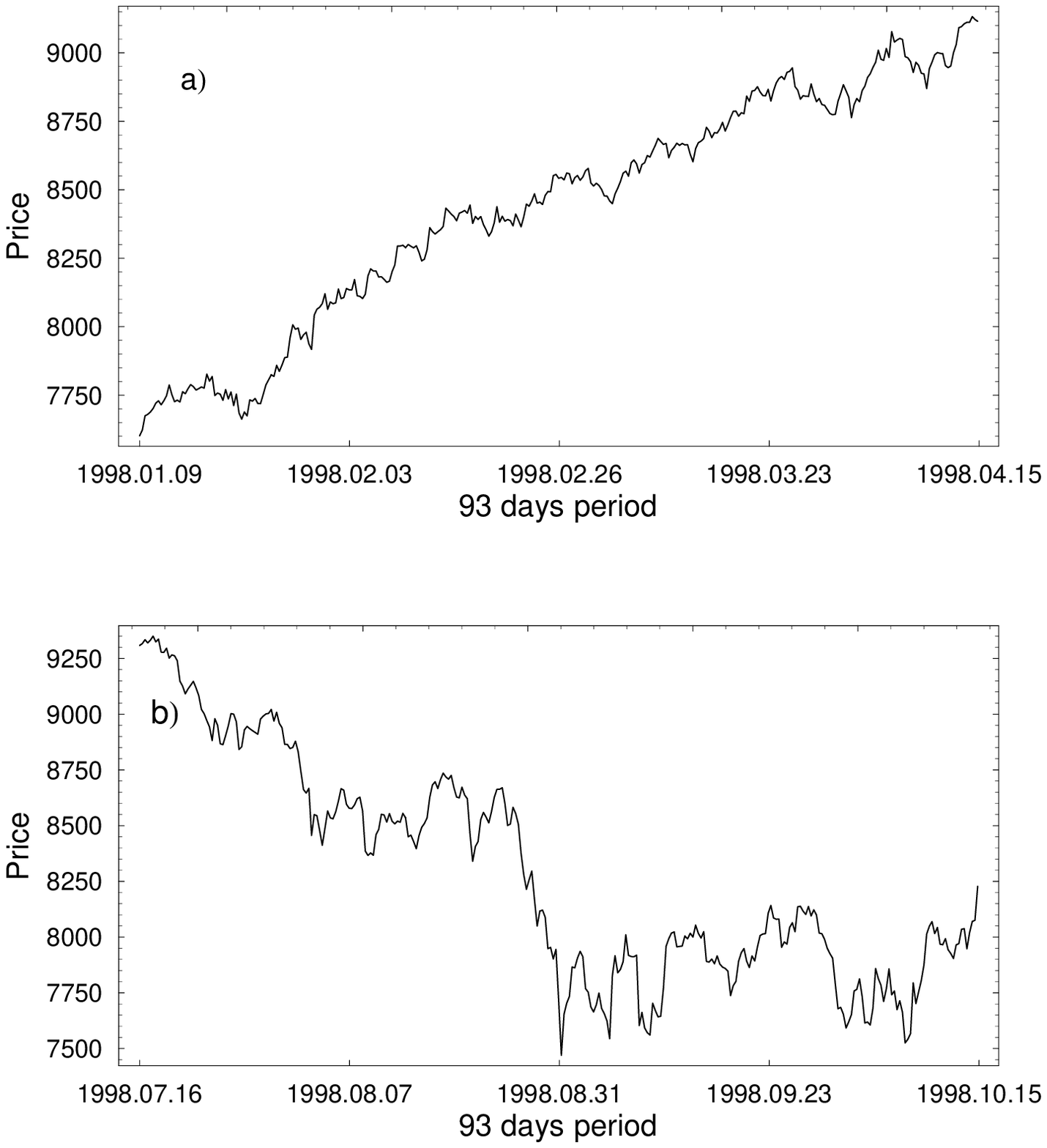}
\begin{footnotesize}
\begin{center}

\begin{tabular}{|c|c|c|c|c|c|c|c|} \hline \bf Data& $\bf P_{11}$& $\bf P_{1-1}$&
$\bf P_{10}$& $\bf P_{-1-1}$& $\bf P_{-11}$& $\bf P_{-10}$& \bf{Hurst exp.}\\
\hline \bf DJIA(increase) &0.534 &0.415  &0.051  &0.523   &0.424  &0.052 &0.495\\
\hline \bf DJIA(decrease) &0.609  &0.363  &0.028  &0.629  &0.342
&0.029 &0.505\\ \hline
\end{tabular}

\end{center}
\end{footnotesize}
\caption{DJIA during its long-term global increase in the period
09.01.1998 - 15.04.1998 (a) and during its long-term global
decrease in the period 16.07.1998 - 15.10.1998 (b). These both
periods correspond to the same number (93) of trading days. Tab.3.
Conditional probabilities $p_{\alpha, \beta}$ and the Hurst
exponents for the DJIA $\Delta t=1$ min changes corresponding to
the time periods as in Fig.4.a (DJIA /increase) and as in Fig.4.b
(DJIA/decrease), respectively.}
\end{figure}

\section{Conclusion}

The above observations are intriguing and of course demand a much
more systematic study as they carry a potential to shed more light
on mechanism of the stock market dynamics. The present study
however already indicates direction concerning this specific
issue. There definitely exist higher order correlations in the
financial dynamics that escape detection within the conventional
methods. In this connection the wavelet based formalism, due to
its ability to focus on local effects, seems to offer a promising
frame to develop consistent related methodology, such that a link
to multifractality perhaps can also be traced. The wavelet based
formalism can also be generalised to account for the asymmetry in
persistence - an effect identified above. Finally and ultimately,
one needs to develop a realistic theoretical model of the
financial dynamics such that also the above effects can be
incorporated. A variant of the generalised Weierstrass random walk
as developed by Kutner~\cite{Kut} may appear an appropriate
solution especially that the Weierstrass-type functions may
incorporate log-periodicaly a kind correlations that underlay the
financial dynamics ~\cite{Dro3,Dro4}.

\end{document}